\documentclass[12pt]{article}
\def\pslash{\hbox{/\kern-.5800em$p$}}

\def\gappeq{\mathrel{\rlap {\raise.5ex\hbox{$>$}}
{\lower.5ex\hbox{$\sim$}}}}
\def\lappeq{\mathrel{\rlap{\raise.5ex\hbox{$<$}}
{\lower.5ex\hbox{$\sim$}}}}

\def\gappeq{\mathrel{\rlap {\raise.5ex\hbox{$>$}}
{\lower.5ex\hbox{$\sim$}}}}

\def\lappeq{\mathrel{\rlap{\raise.5ex\hbox{$<$}}
{\lower.5ex\hbox{$\sim$}}}}

\begin{document}
\pagestyle{empty}
\begin{flushright}
UMN-TH-2325/04\\
November 2004
\end{flushright}
\vspace*{5mm}

\begin{center}
{\Large\bf Partly Supersymmetric Grand Unification}
\vspace{1.0cm}

{\sc Tony Gherghetta}\\
\vspace{.5cm}
{\it\small {School of Physics and Astronomy\\
University of Minnesota\\
Minneapolis, MN 55455, USA}}\\
\end{center}

\vspace{1cm}
\begin{abstract}

It is shown how grand unification can occur in models which are partly 
supersymmetric. The particle states which are composite do not 
contribute to the running of gauge couplings above the compositeness 
scale, while the elementary states contribute the usual large logarithmns. 
This introduces a new differential running contribution to the gauge 
couplings from partly composite $SU(5)$ matter multiplets. In particular,
for partly supersymmetric models, the incomplete $SU(5)$ elementary matter 
multiplets restore gauge coupling unification even though the usual 
elementary gaugino and Higgsino contributions need not be present.

\end{abstract}

\vfill
\begin{flushleft}
\end{flushleft}
\eject
\pagestyle{empty}
\setcounter{page}{1}
\setcounter{footnote}{0}
\pagestyle{plain}


\section{Introduction}

The impressive unification of the gauge couplings in the minimal 
supersymmetric standard model (MSSM) is touted as indirect evidence for 
supersymmetry at low energies. This unification results from the superpartner 
contribution of incomplete $SU(5)$ mulitplets, namely, the gauginos 
and Higgsinos. As a bonus these sparticles also provide a suitable dark matter 
candidate. However, the supersymmetric model generically leads to flavour, CP and 
gravitino/moduli problems. While elaborate mechanisms do exist that solve these problems the 
generic problem arises because the scale of supersymmetry breaking is low. It is 
tempting then to consider natural models of high-scale supersymmetry breaking.

Recently, a partly supersymmetric model~\cite{gp4} was constructed
in which only the Higgs sector is supersymmetric. This model naturally 
arises in the five dimensional warped geometry~\cite{rs}, but has a purely 
four dimensional interpretation as a partly composite model. Both the Higgs 
and Higgsino are composite states, with the Higgsino playing the role of the 
dark matter candidate. All superpartners in this model (except, possibly the 
third generation sparticles), receive large supersymmetry breaking masses, 
thereby providing a simple solution to the supersymmetric flavour, CP and 
gravitino/moduli problems (where the gravitational sector problems 
require considering local supersymmetry in the bulk~\cite{luty}).

However the success of the supersymmetric standard model is also measured 
in the precise unification of the gauge couplings. The dominant effect 
in the differential running comes from the gaugino and Higgsino contributions.
But in the partly supersymmetric model the required gaugino contributions 
are not present, since these states are decoupled at low energies, and
the Higgsinos (and Higgs) are composite and consequently do not contribute to
the running above the TeV compositeness scale. Thus, it would seem that 
the impressive gauge coupling unification in the MSSM is lost.

In this paper we show that the differential running 
effects of the gauginos and Higgsinos are replaced in the partly 
supersymmetric model by those arising from part of the standard model 
fermions. This is because fermions in $SU(5)$ representations will 
be partly composite. This will be natural for the third generation fermions 
which have large Yukawa couplings and therefore need a large
wavefunction overlap with the composite Higgs. Moreover, a composite
top quark is required to trigger electroweak symmetry breaking~\cite{gp4}.
Assuming that an $SU(5)$ GUT symmetry is broken by boundary conditions 
requires that the standard model fermions form part of
{\it different} $SU(5)$ multiplets. Then, since only part of the 
third generation is composite, a differential effect is introduced in
the running of the gauge couplings because the composite fermions
do not contribute to the running above the compositeness scale. 
This differential effect in the gauge coupling running from fermions 
can help to restore the unification of the couplings at a high energy 
scale that was lost from the gauginos and Higgsinos.

In the slice of AdS$_5$ there is a large separation between the
infrared (IR) and ultraviolet (UV) energy scales because these theories are 
holographically dual to a four-dimensional CFT~\cite{adscft,phenoadscft}.
Remarkably, this leads to logarithmic running in these models~\cite{log}, 
although the issue of gauge coupling unification is more subtle. The running 
above the IR (or TeV) scale $T\equiv k e^{-\pi k R}$ must be understood 
in terms of the Planck brane-brane correlator, and since gauge bosons live 
in the bulk this cannot be done with the zero mode correlator~\cite{gr}. 
Nonetheless, at low energies, $p\leq T$, the predictions are identical 
because the Kaluza-Klein states do not contribute to the running, and so
either correlators can be used. This is the basis for studying gauge 
coupling unification in warped models. In previous work, large but natural 
threshold corrections were used to restore grand unification 
in nonsupersymmetric models~\cite{ads}. 
However in warped supersymmetric models~\cite{gp2} the conventional MSSM 
unification can be obtained~\cite{wgu}, albeit with the usual TeV mass scale 
superpartners. In fact this class of models is identical to the usual MSSM
coupled to a strong CFT sector which is responsible for dynamical 
supersymmetry breaking. Instead we will consider gauge coupling
unification in natural high-scale supersymmetric models where, except 
for sparticles associated with the Higgs sector, the superpartners 
are superheavy and decoupled from low energies. In fact the idea of 
having part elementary matter fields contributing to the differential
running of gauge couplings is quite general and we will also briefly mention
other partly supersymmetric possibilities (although see Ref.~\cite{acs} for 
a nonsupersymmetric possibility).

\section{GUT breaking on the UV brane}

The geometrical setting of the model is a slice of AdS$_5$ where the 
fifth dimension is compactified on an orbifold $S^1/Z_2$ of radius $R$ 
with the metric
\begin{equation}
   ds^2 = e^{-2 k |y|}\eta_{\mu\nu} dx^\mu dx^\nu + dy^2~,
\end{equation}
where $0 \leq y\leq \pi R$, $k$ is the AdS curvature scale, and 
the Minkowski metric $\eta_{\mu\nu}$ has signature $(-+++)$. At the 
orbifold fixed points $y^\ast=0$ and $y^\ast=\pi R$ there are two 
3-branes, the UV (or Planck) brane and IR (or TeV) brane, respectively. 
Generically fermions are introduced into the five-dimensional
bulk as Dirac fermions with masses  $m_\Psi = c k \epsilon(y)$, where 
$\epsilon(y)=y/|y|$, and $c$ is a dimensionless parameter. The 
zero modes of the bulk fermions are identified with the usual low 
energy standard model fermions. If $c>1/2~(c<1/2)$ then the fermion 
zero mode is localised towards the Planck (TeV) brane~\cite{gp1}.

We will assume that there is an $SU(5)$ gauge symmetry in the bulk that
is broken on the UV brane by boundary conditions. This occurs by imposing
Neumann $(+)$ or Dirichlet $(-)$ conditions on the bulk fields corresponding
to even or odd reflections, respectively, about the orbifold fixed points.
The $SU(5)$ gauge bosons form an ${\cal N}=2$ vector multiplet 
${\cal V}=(V,S)$, where $V$ is an ${\cal N}=1$ vector multiplet and 
$S$ is an ${\cal N}=1$ chiral multiplet 
with the boundary conditions
\begin{equation}
  V=\left[\begin{tabular}{c}
            $V_\mu^a(+,+)$\\ 
            $V_\mu^A(-,+)$
            \end{tabular}\right]~,\qquad\qquad
  S=\left[\begin{tabular}{c}
            $S^a(-,-)$\\ 
            $S^A(+,-)$
            \end{tabular}\right]~,
\end{equation}
and the indices $a (A)$ run over the unbroken (broken) generators.
In particular notice that the $SU(5)$ symmetry remains unbroken on 
the TeV-brane, and the only zero modes in the spectrum are identified 
with the $SU(3)\times SU(2) \times U(1)_Y$ gauge fields. At the nonzero
mode level, the $X,Y$ gauge bosons and the $SU(5)$ adjoint scalar states 
obtain TeV-scale masses.

Similarly, the Higgs sector is supersymmetric and 
contains two Higgs doublets which are
embedded into two ${\cal N}=2$ bulk hypermultiplets, ${\cal H}=(H_5,H^c_5)$
and $\bar {\cal H}=(\bar H_5,\bar H^c_5)$, each transforming
in the ${\bf 5}$ of $SU(5)$. The boundary conditions are
\begin{equation}
  H_5 =\left[\begin{tabular}{c}
            $H_2(+,+)$\\ 
            $H_3(-,+)$
            \end{tabular}\right]~, \qquad\qquad
  H^c_5 =\left[\begin{tabular}{c}
            $H_2^c(-,-)$\\ 
            $H_3^c(+,-)$
            \end{tabular}\right]~,
\end{equation}
and similarly for $\bar {\cal H}$. Thus the only zero modes are the
two Higgs doublets, $H_2(+,+)$ and ${\bar H}_2(+,+)$, as in the MSSM.
This choice of boundary conditions neatly solves the doublet-triplet 
splitting problem~\cite{kawamura}.
The fields with boundary conditions $(\mp,\pm)$ and $c<-1/2 (c>1/2)$
obtain masses of order
\begin{equation}
    m_{(\mp,\pm)} \simeq 2\sqrt{c^2-\frac{1}{4}} 
        k~e^{(\pm c-\frac{1}{2})\pi k R}~,
\end{equation}
while for $c> -1/2 (c<1/2)$ their masses are 
\begin{equation}
    m_{(\mp,\pm)} \simeq k~e^{-\pi k R}~.
\end{equation}

The standard model matter fields are also embedded into 
bulk ${\cal N}=2$ hypermultiplets. 
However, while it would seem natural to put all the fermions of a 
single generation into a single ${\bf 5}$ and ${\bf 10}$, parity assignments 
actually require the quarks and leptons to arise from different $SU(5)$ bulk
hypermultiplets~\cite{hn1,hmr,bhn}. Not only does this mechanism elegantly
explain why the fermions need not satisfy the $SU(5)$ mass relations, but
we will see that this fact is responsible for the differential running 
of the gauge couplings~\cite{ads}. For each generation we will suppose 
that there are bulk hypermultiplets
$({\bf 5}_1,{\bf 5}_1^c)+({\bf 5}_2,{\bf 5}_2^c)$ and
$({\bf 10}_1,{\bf 10}_1^c)+({\bf 10}_2,{\bf 10}_2^c)$ with boundary conditions
\begin{eqnarray}
  &&{\bf 5}_1 = L_1(+,+) + d_1^c(-,+)~,\quad
    {\bf 5}_1^c = L_1^c(-,-) + d_1(+,-)~, \\
  &&{\bf 5}_2 = L_2(-,+) + d_2^c(+,+)~,\quad
    {\bf 5}_2^c = L_2^c(+,-) + d_2(-,-)~,\\
  &&{\bf 10}_1 = Q_1(+,+) + u_1^c(-,+) +e_1^c(-,+)~,\\
  &&{\bf 10}_1^c = Q_1^c(-,-) + u_1(+,-) +e_1(+,-)~,\\
  &&{\bf 10}_2 = Q_2(-,+) + u_2^c(+,+) + e_2^c(+,+)~,\\
  &&{\bf 10}_2^c = Q_2^c(+,-) + u_2(-,-) + e_2(-,-)~,
\end{eqnarray}
where the standard model fermions are identified with the zero modes of
the fields with (+,+) boundary conditions. Notice that even though each 
standard model generation arises from different ${\bf 5} + {\bf 10}$ fields, 
the usual charge quantization and hypercharge assignments arising from 
a single ${\bf 5} + {\bf 10}$ are still explained~\cite{hn1,hmr}. This feature
also explains why tree-level proton decay is not a problem in these models. 
There is simply no coupling between $X,Y$ gauge bosons and standard model 
fields $L$ and $d^c$ since these fields arise from different $SU(5)$ 
multiplets. This is also true for couplings between standard model particles 
and the coloured Higgs triplets. However, a bulk $U(1)$ symmetry must be 
introduced in order to prevent proton decay from higher-dimensional 
operators~\cite{wgu,ads}.

Now in the partly supersymmetric standard model, supersymmetry is 
assumed to be broken on the UV (or Planck) brane. This means that fields which
are coupled to the UV brane directly feel the high-scale supersymmetry 
breaking and decouple from the low energy spectrum. In particular, 
the gauginos of the ${\cal N}=1$ vector multiplets receive Planck scale 
masses. For the bulk hypermultiplets only superpartner fields with 
$c\gappeq 1/2$ receive Planck scale masses, while hypermultiplets with 
$c\lappeq -1/2$ are insensitive to the UV supersymmetry breaking. Therefore 
the zero modes of bulk fields with $c\lappeq -1/2$ are the only particles 
in the supersymmetric sector. In the dual 4D theory these states are also 
seen to be composite states of the CFT, which explains why they are not 
sensitive to the UV supersymmetry breaking~\cite{gp4}.

In particular the bulk Higgs hypermultiplets are assumed to have 
$c_H\lappeq -1/2$, which means that the Higgs is localised on the 
IR brane, and the Higgs sector is supersymmetric (where for simplicity
$c_H=c_{\bar H}$). In the limit that 
$c_H \rightarrow -\infty$, there are two localised Higgs doublets, as well 
as two Higgs triplets on the TeV brane. This is consistent with the fact 
that the $SU(5)$ symmetry is only broken on the Planck brane.
Since the Higgs is localised on the TeV brane, the hierarchies in the 
Yukawa couplings of the first two generations can be generated by localising 
these fermions towards the Planck brane, and therefore the first two 
generations are elementary. However, to generate the larger third generation 
fermion masses requires some fermions to necessarily be localised towards 
the TeV brane, and consequently are composite. A composite top quark
is also crucial to generate radiative electroweak symmetry 
breaking~\cite{gp4}. Thus, the third generation fermion sector is naturally 
part elementary and part composite. 

Specifically, consider the case where $c_{{\bf 5}_1},c_{{\bf 10}_1} \gappeq 
1/2$, while $c_{{\bf 5}_2},c_{{\bf 10}_2} \lappeq -1/2$. This means that 
$Q_1(+,+)$ and $L_1(+,+)$ are localised towards the Planck brane, and are 
elementary fields in the 4D dual description, 
while $u_2^c(+,+)$, $d_2^c(+,+)$ and $e_2^c(+,+)$ are localised towards 
the TeV brane and are composite states. The crucial point is that since 
the zero modes with $c\lappeq -1/2$ are composite then they do not contribute 
to the running above the TeV scale, while the elementary states do. Thus, 
above the TeV scale the elementary standard model fermions of the third 
generation form an incomplete $SU(5)$ multiplet and contribute to the 
differential running of the gauge couplings!

Actually this is not the entire story because
there are other elementary fermion contributions that arise from
the bulk hypermultiplet fields with $(\pm,\mp)$ boundary conditions. 
Naively, these contributions arise from the fact that there should be 
a source field for each bulk Dirac fermion component. However, from the 
bulk point of view one cannot simultaneously introduce two Weyl fermion 
sources on the boundary because the 5D Lagrangian only contains first-order 
derivatives and the boundary value of one Weyl fermion always vanishes since 
it is odd. Nonetheless for a bulk Dirac fermion with $c\lappeq -1/2$ and 
$(-,+)$ boundary conditions, that is described in the CFT 
with a left-handed source, the 
massless composite fermion zero mode $\psi_L^{(0)}$ must decouple because 
there can be no massless states. Since there is no coupling to the 
left-handed source a new elementary state $\chi_R^{(0)}$ must appear 
which pairs up with $\psi_L^{(0)}$ to form a massive Dirac state~\cite{cp}. 
Consequently, since the new right-handed state is elementary it will 
also contribute to the running above the TeV scale. 

So, for $c\lappeq -1/2$, consider the ${\cal N}=2$ hypermultiplet pairs 
$(Q_2(-,+),Q_2^c(+,-))$ and $(L_2(-,+),L_2^c(+,-))$. The lowest lying 
fermion will be a Dirac state which consists of a composite Weyl fermion
from the $(-,+)$ field married to an elementary Weyl fermion from the
$(+,-)$ field. The elementary fermion will thus contribute to the running. 
Note that by supersymmetry there is also an elementary complex scalar, but
it will receive a large supersymmetry breaking mass from the UV brane, and
therefore does not contribute to the low energy running of the gauge couplings.
The remaining ${\cal N}=2$ hypermultiplets of the $SU(5)$ gauge multiplet with 
$c\lappeq -1/2$ are $(d_2^c(+,+),d_2(-,-))$, $(u_2^c(+,+),u_2(-,-))$ and 
$(e_2^c(+,+),e_2(-,-))$.
In this case the corresponding zero mode fields are all composite and do not 
contribute to the running above the TeV (or compositeness) scale. 
Thus, with respect to the usual MSSM contribution to the gauge coupling
running this means that we subtract the contribution of $u_2^c,d_2^c$ 
and $e_2^c$, and include an extra fermion contribution from $Q_2^c$ and 
$L_2^c$ to the running. For the differential running of the gauge couplings
this is effectively equivalent to subtracting twice the contribution of 
$u_2^c,d_2^c$ and $e_2^c$, since complete $SU(5)$ multiplets do not 
change the differential running.

\section{Analysis of gauge coupling unification}

Successful gauge coupling unification relies on all three gauge couplings
being within a few percent of the low energy observed values. In the setup
we are considering here, with the bulk gauge symmetry broken
by boundary conditions, the low energy gauge couplings, $g_a$ are given by
\begin{equation}
\label{gceqn}
     \frac{1}{g_a^2(p)} = \frac{\pi R}{g_5^2} + \frac{1}{g_{Ba}^2(\mu)} 
     + \frac{1}{8\pi^2}\Delta_a(p,\mu)~,
\end{equation}
where $\mu$ is a subtraction scale, $g_{Ba}$ are boundary couplings, 
and $\Delta_a$ are the one-loop corrections. 
The first term in (\ref{gceqn}) is the contribution from 
the tree-level gauge coupling $g_5$ in the bulk. This contribution is universal
because SU(5) is not broken in the bulk. 
Since $SU(5)$ is broken on the UV brane the UV boundary coupling encodes 
unknown $SU(5)$-violating contributions. This would seem to imply that 
there are no predictions for the low energy couplings. However, using NDA 
it can be shown that the $SU(5)$-violating contributions are sub-dominant 
compared to the one-loop contributions $\Delta_a$~\cite{nda}.
In fact this is ensured by choosing $\mu \simeq \Lambda$, where $\Lambda$
is the scale where the theory is effectively strongly coupled~\cite{bc,gr2}. 
In this way the boundary coupling contains no large logarithmns since
$g_{Ba}(\Lambda)\simeq 4\pi$.
Thus, given that the universal tree-level coupling $\pi R/g_5^2\simeq 
{\cal O}(1)$ the successful prediction of the low energy couplings is not 
lost. The dominant contribution will arise from the logarithmically 
enhanced terms of $\Delta_a$.

The large logarithmns that arise in $\Delta_a$ can be calculated by using 
the bulk zero-mode Green functions~\cite{choi}. However, these expressions 
cannot be extrapolated beyond the TeV scale, since the effective field theory 
breaks down. Instead, we can interpret the large logarithmns as arising from 
the running of the Planck brane-brane correlator down to low 
energy~\cite{gr2}. 
Of course, at energies below the TeV scale the zero-mode Green functions 
and Planck correlators will agree at the leading log level, but only the 
Planck brane-brane correlator can make sense all the way up to high energies.
At the leading log level the generic form of $\Delta^a$ is
\begin{equation}
     \Delta^a(p,\Lambda) = b^a\log\frac{\Lambda}{p} + SU(5)~{\rm universal}~,
\end{equation}
where $b^a$ are the one-loop $\beta$-function coefficients of the particles
that contribute to the running. Since we can safely neglect the 
non-universal boundary coupling contributions in Eq.(\ref{gceqn}), we can 
eliminate the universal piece and the scale $\Lambda$ to obtain the
low energy prediction at $p=M_Z$, namely
\begin{equation}
\label{Bexpt}
    B=\frac{b_3-b_2}{b_2-b_1}= 0.717\pm0.008~,
\end{equation}
where the error is due to the experimental determination of the low energy
couplings. In the 4D CFT picture the one-loop bulk zero mode Greens 
functions in Ref.~\cite{choi} can be interpreted as two-loop contributions 
involving CFT states. Thus, by expanding the bulk expressions in $p/T$ we 
can obtain some of the NLO corrections in the CFT.

\subsection{Gauge sector}

Let us now calculate the gauge coupling contributions in the partly
supersymmetric model. In the gauge sector the zero-mode contribution from the
unbroken gauge group $SU(3)\times SU(2)\times U(1)$ only arises from
the gauge bosons, since the gauginos have decoupled below the Planck scale.
Using the expressions in Ref.~\cite{choi}
we obtain for ${\cal V}^a_{++} = (V_\mu^a(+,+), S^a(-,-))$
\begin{equation}
\label{delVpp}
     \Delta^a({\cal V}_{++})=T_a({\cal V}_{++}) \left[ -\frac{11}{3}
      \ln \frac{k}{p}+\frac{3}{2} \pi k R +\ln 2kR \right]~,
\end{equation}
where $p\lappeq T$ and $T_a({\cal V}_{++})$ is the Dynkin index. For 
$SU(3)\times SU(2)\times U(1)$ we have $T_a({\cal V}_{++}) =(0,2,3)$ 
and only the zero mode gauge fields contribute a large logarithmn, which 
can be thought of as logarithmic running from momenta $p\lappeq T$ up to the 
scale $k$. The $\pi kR (= \ln \frac{k}{T})$ contribution in (\ref{delVpp})
arises from nonanalytic operators in the bulk like 
$\sqrt{\cal R} FF$~\cite{bc}. 
In the holographic picture this corresponds to the calculable part of 
the NLO CFT correction. We will see that since the bulk preserves the 
$SU(5)$ gauge symmetry the total contributions of this form will be universal.
For the broken generators, we find that for $p\lappeq T$ and 
${\cal V}^A_{-+} = (V_\mu^A(-,+), S^A(+,-))$
\begin{equation}
     \Delta^a({\cal V}_{-+})=T_a({\cal V}_{-+})\left[
      \frac{3}{2} \pi k R + \ln\frac{2}{\pi}\right]~.
\end{equation}
In particular notice that there are no large logarithmn contributions 
consistent with the fact that there are no elementary fields in 
${\cal V}^A_{-+}$. There is also a $\pi k R$ contribution arising from the
nonanalytic bulk operators. Using the fact that $T_a({\cal V}_{-+}) =(5,3,2)$
the total contribution from the gauge supermultiplet $\cal V$ is
\begin{equation}
\label{delV}
     \Delta^a({\cal V})=b^a_{\cal V}\left[\ln \frac{k}{p}
               -\frac{1}{3}\ln(\pi kR)\right]+\frac{15}{2} \pi k R 
        +5\ln\frac{2}{\pi}~,
\end{equation}
where 
\begin{equation}
\label{bvec}
      b^a_{\cal V} =(0, -\frac{22}{3}, -11)~.
\end{equation}
Notice that the first term in (\ref{delV}) represents the contribution
from the elementary gauge boson zero modes. There is no corresponding 
contribution from the gaugino zero-modes because they have decoupled
at low energies due to the high-scale supersymmetry breaking. In addition
we have neglected the threshold effects arising from supersymmetry breaking,
which should be comparable to those in the MSSM.

The second term in (\ref{delV}) represents a two-loop effect in the CFT 
which is not negligible because the source coupling to the current operator 
in the CFT is marginal. In the CFT picture the term $\ln(\pi k R) = 
\ln\ln\frac{k}{T}$ can be written as a double logarithmn. To see this term
as a two-loop effect consider, in particular, a one-loop gauge boson diagram 
where one of the gauge boson propagators is corrected by the CFT. 
By using the expression for the corrected propagator in Ref.~\cite{log}, 
the amplitude of this two-loop CFT diagram then becomes 
\begin{equation}
    \int d^4p~\frac{1}{p^2 \ln\frac{p}{k}}~\frac{1}{p^2} \sim 
       \ln\ln\frac{p}{k}~,
\end{equation}
which for $p\simeq T$ has the appropriate form.
Furthermore there are other two-loop contributions arising
from 3-point and 4-point vertex corrections. By invoking Ward identities 
these contributions are related to the above corrected gauge boson propagator
contribution. Thus, the $\ln(\pi kR)$ term in (\ref{delV})
represents the total from each of these contributions.
Also as expected we see that the $\pi k R$ contribution in (\ref{delV}) 
is universal.

\subsection{Higgs sector}

In the Higgs sector ${\cal H}_{++}= (H_2(+,+),H_2^c(-,-))$ 
(and similarly for $\bar{\cal H}_{++}$),
with $c_H\lappeq -1/2$ there are no large logarithmic 
contributions from the zero-mode doublets 
because all these fields are composite. However since the expression
$\Delta^a$ is for $p\lappeq T$ there is 
a small contribution arising from the running below
the scale $T$ from both the Higgs and Higgsino, given by
\begin{equation}
\label{delHpp}
     \Delta^a({\cal H}_{++})=T_a({\cal H}_{++}) \left[
      \ln \frac{T}{p}+c_H \pi k R +\ln\pi(\frac{1}{2}-c_H)\right]~,
\end{equation}
where $T_a({\cal H}_{++})=(3/10,1/2,0)$.
This is consistent with the fact that for $c_H\lappeq -1/2$ the zero mode
is composite and does not contribute to the running above the
compositeness scale $T$. There are also terms arising from bulk operators,
but again these will be universal. In the presence of supersymmetry breaking 
on the UV brane there are mass threshold effects for the Higgs and Higgsino
which are not included in (\ref{delHpp}). Since we are mainly interested 
in the large log behaviour of $\Delta^a$ we will not consider these small 
effects further.

However the triplet Higgs
hypermultiplets ${\cal H}_{-+}= (H_3(-,+),H_3^c(+,-))$ 
(and similarly for $\bar{\cal H}_{-+}$),
have $(\mp,\pm)$ boundary conditions and as explained 
earlier the lowest lying fermion state is a massive Dirac state with 
mass $m_{-+}$ that consists of a composite fermion pairing with an 
elementary fermion. There is no running for momenta below $m_{-+}$, but
for $m_{-+}\lappeq p \lappeq T$ we obtain
\begin{equation}
\label{delHmp}
     \Delta^a({\cal H}_{-+})=T_a({\cal H}_{-+}) \left[\frac{2}{3}
      \ln \frac{k}{p} +\ln \frac{T}{p}+c_H\pi k R +
     +\ln2\pi(c_H^2-\frac{1}{4})\right]~,
\end{equation}
and similarly for $\bar{\cal H}$, where $T_a({\cal H}_{-+})=(1/5,0,1/2)$.
The hybrid nature of the Dirac fermion zero-mode can be seen from 
the first two terms in (\ref{delHmp}). In the first term 
only an elementary Weyl fermion contributes a large logarithmn
since the elementary scalar partner receives a large supersymmetry-breaking
mass and decouples from the low energy spectrum. The second term shows 
the contribution from both the scalar and fermion composite triplets whose
running stops at the compositeness scale $T$. Thus, combining (\ref{delHpp}) 
and (\ref{delHmp}), the total Higgs contribution for $c_H\lappeq -1/2$ 
and $m_{-+}\lappeq p \lappeq T$ is
\begin{equation}
\label{delH}
     \Delta^a({\cal H} +\bar{\cal H})=b^a_{{\cal H}+\bar{\cal H}}
       \left[\ln \frac{k}{p} + \ln(-2c_H-1)\right]
       + \ln \frac{T}{p} + c_H\pi k R +\ln\pi(\frac{1}{2}-c_H)~,
\end{equation}
where the two elementary triplet Weyl Higgsinos from both Higgs 
hypermultiplets $\cal H$ and $\bar{\cal H}$ contribute an amount
\begin{equation}
\label{bHiggs}
     b^a_{{\cal H}+\bar {\cal H}}=\left(\frac{4}{15},0,\frac{2}{3}\right)~. 
\end {equation}

\subsection{Matter sector}

For the bulk matter the first two generations always have 
$c_{{\bf 5}_i},c_{{\bf 10}_i}\gappeq 1/2$, and
consequently all the zero modes are elementary and localised towards the 
Planck brane. The scalar zero modes (squarks and sleptons) will decouple
at low energies because they feel the high-scale supersymmetry breaking,
and so only the zero-mode fermions will give rise to a large logarithmic
contribution. Thus, for a standard model fermion generation $I$ with
$c\gappeq 1/2$ one obtains
\begin{eqnarray}
\label{del12}
     \Delta^a({\bf 5}_i^{(I)}+{\bf 10}_i^{(I)})&=&\frac{4}{3} \ln \frac{k}{p}
      +T_a({\bf 5}_{i++})\ln(c_{{\bf 5}_i}-\frac{1}{2})
      +T_a({\bf 10}_{i++})\ln(c_{{\bf 10}_i}-\frac{1}{2})\nonumber\\
      &&\qquad\quad  -\frac{1}{2}(c_{{\bf 5}_i}+3c_{{\bf 10}_i})\pi k R~,
\end{eqnarray}
where $i=1,2$ and $T_a({\bf 5}_{i++}) (T_a({\bf 10}_{i++}))$ is the
Dynkin index of the $(+,+)$ fields in ${\bf 5}_i ({\bf 10}_i)$.
The first term in (\ref{del12}) is the usual $SU(5)$ universal 
contribution from one generation of standard model fermions. The 
second and third terms are higher-order nonuniversal contributions,
while the last term is another universal contribution.

On the other hand the third generation is assumed to be 
partly composite. Only the zero mode fermions in $L_1(+,+)$ and $Q_1(+,+)$ 
will contribute to the running since $b_2^c(+,+)$, $t_2^c(+,+)$ and 
$\tau_2^c(+,+)$ are composite. In addition the hypermultiplets $L_2$ and $Q_2$
with $c_{{\bf 5}_2},c_{{\bf 10}_2}\lappeq -1/2$ and $(\mp,\pm)$ boundary 
conditions have zero-mode Dirac fermion states. The Dirac fermion state again 
contains an elementary Weyl fermion which contributes a large logarithmn to 
the running. In this case we find
\begin{eqnarray}
\label{del3}
     \Delta^a({\bf 5}_i^{(3)}+{\bf 10}_i^{(3)})&=&b^a_{(3)}\ln \frac{k}{p}
     +\frac{4}{3}\ln \frac{T}{p}
     +\frac{1}{2}(c_{{\bf 5}_i}+3c_{{\bf 10}_i})\pi k R\nonumber\\
   && +\ln(\frac{1}{2}-c_{{\bf 5}_2})+3\ln(\frac{1}{2}-c_{{\bf 10}_2})
    \nonumber\\
   && +T_a({\bf 5}_{1++})\ln(c_{{\bf 5}_1}-\frac{1}{2})(-c_{{\bf 5}_2}-\frac{1}{2})
    \nonumber\\
   && +T_a({\bf 10}_{1++})\ln(c_{{\bf 10}_1}-\frac{1}{2})(-c_{{\bf 10}_2}
    -\frac{1}{2})~,
\end{eqnarray}
where 
\begin{equation}
\label{bmatter}
        b^a_{(3)}=\left(\frac{8}{15}, \frac{8}{3}, \frac{4}{3}\right)~.
\end {equation}
In (\ref{del3}) we clearly see the differential running effect in the 
first term arising from the elementary fermion zero modes. The second term
represents the composite fermion zero modes which stop running
at the compositeness scale $T$, with the expected $\beta$-function 
coefficient from the ${\bf 5}$ and ${\bf 10}$ 
representations of $SU(5)$. Note that we have assumed that the 
corresponding scalar partners receive masses at the scale $T$ and
therefore their contribution is not included in the second term.

\subsection{The total contribution}

If we now add up all the $\Delta^a$ contributions (\ref{delV}), (\ref{delH}), 
(\ref{del12}) and (\ref{del3}) arising from the elementary states 
in the model then at the leading log level we obtain the total contribution
\begin{equation}
     \Delta^a = b^a_{total} \ln\frac{k}{p} + SU(5)~{\rm universal}~,
\end{equation}
where
\begin{equation}
       b^a_{total}=\left(\frac{52}{15},-2,-\frac{19}{3}\right)~.
\end {equation}
These values of the $\beta$-function coefficients gives $B=0.793$ which 
is approximately within $10\%$ of the experimental value (\ref{Bexpt}).
This compares with the Standard Model with no Higgs contribution 
(as in the original RS1 model) which gives $B=0.5$, a $40\%$ discrepancy.
Remarkably, the contribution from the partly composite 
third generation fermion sector has restored the low energy prediction to
a level that can be realistically explained by threshold and higher-loop
effects. The unification scale occurs at $\simeq 10^{16}$ GeV, which is 
similar to the usual MSSM, and the value of the unified coupling
$\alpha_{GUT} \sim {\cal O}(1)$.

\section{Other possible models}

We argued earlier that it is natural to have the third generation be
part elementary and part composite. This was primarily motivated by 
the fact that we would like the extra dimensions to explain the 
Yukawa coupling hierarchies. However if we relax this condition
and introduce small parameters for the Yukawa couplings, as is usually 
done in the supersymmetric standard model, then there are many more 
possibilities for which fermions can be composite. In fact, one can
achieve unification at the leading log level that is just as good
as the MSSM. Of course we have implicitly assumed all along that we 
can arbitrarily vary the bulk mass parameter $c$.
However precision electroweak data can limit the range
of allowed $c$ parameters if the Kaluza-Klein mass scale is
less than several TeV~\cite{adms}. In order to explore the complete
parameter space of models with partly supersymmetric grand unification 
we will assume that the Kaluza-Klein mass scale is heavy enough so that 
$c$ is not constrained. This means that in these models 
there is a little hierarchy problem as in the MSSM.

To explore the most possibilities we need to generalise 
the parity assignments of the fields. In the simple $Z_2$ orbifold where 
the boundary conditions were chosen 
to break $SU(5)$ down to the standard model gauge group on the UV brane
the parity assignments were restricted. In the ${\bf 10}$ representation
the $u^c$ and $e^c$ parity assignments are always opposite those of $Q$. 
However, one can always introduce extra fermion states on the boundaries that 
marry the unwanted zero modes so that only one field in each ${\bf 10}$ has
$(+,+)$ boundary conditions. Thus we can generalise to three ${\bf 10}$ 
representations for each generation with parity assignments
\begin{eqnarray}
  &&{\bf 10}_1 = Q_1(+,+) + t_1^c(-,+) + \tau_1^c(-,+)~,\\
  &&{\bf 10}_2 = Q_2(-,+) + t_2^c(+,+) + \tau_2^c(-,+)~,\\
  &&{\bf 10}_3 = Q_3(-,+) + t_3^c(-,+) + \tau_3^c(+,+)~,
\end{eqnarray}
where ${\bf 10}_i^c$ have the opposite parity assignments. This then enables 
one to introduce the bulk mass parameter $c_{{\bf 10}_i}$ for each $(+,+)$
field in the ${\bf 10}$. In this way we can always achieve any combination of 
part elementary and part composite zero modes. If we first just
restrict to the third generation then we obtain two other solutions that
are within approximately $10\%$ of the experimental value. So assuming that
the gauge and Higgs sector are exactly as before, the first solution has
only $c_{{\bf 5}_2}^{(3)}, c_{{\bf 10}_2}^{(3)} \lappeq -1/2$ in the matter 
sector corresponding to $t_R$ and $b_R$ composite. The $\beta$-function 
coefficient is $b= (64/15,-2,-19/3)$ leading to a value $B=0.691$ and 
a unification scale of $\approx 10^{15}$ GeV. The second solution has
$c_{{\bf 10}_2}^{(3)}, c_{{\bf 10}_3}^{(3)} \lappeq -1/2$, corresponding to
$t_R$ and $\tau_R$ composite. In this case $b= (22/5,-4/3,-5)$
with $B=0.639$, and the unification scale is near $10^{16}$ GeV. 
In both cases threshold effects and higher-loop effects are needed to 
account for the $10\%$ difference with the experimental value.

Finally we consider more exotic possibilities by allowing fermions in the 
first two generations to be composite. A scan of all possible combinations of 
composite and elementary fermions gives many possibilities that are
within $10\%$ of the experimental value for $B$. Of course not 
all of these solutions are theoretically compelling. However one feature
that follows from allowing part of the first or second generation to be 
composite is that the unification scale can be near the Planck scale. 
An example is $c_{{\bf 10}_2}^{(3)}, c_{{\bf 10}_3}^{(3)}, 
c_{{\bf 10}_3}^{(2)} \lappeq -1/2$, corresponding to $t_R, \tau_R$ and 
$\mu_R$ composite with $b= (23/5,-1/3,-4)$. The unification scale is
interestingly near $10^{18}$ GeV and $B=0.743$ is already quite precise 
at the leading log level. A second representative example has 
$c_R,s_R,t_L,t_R,b_R$ composite with $c_{{\bf 5}_2}^{(2)(3)},
c_{{\bf 10}_2}^{(2)(3)},c_{{\bf 10}_1}^{(3)}\lappeq -1/2$. In this case 
$b= (77/15,-5/3,-20/3)$ with $B=0.735$, but the unification scale is 
lower $\approx 10^{13}$ GeV. Of course threshold and higher-loop effects 
are not included in these examples. These are only a small sample of the 
many permutations of elementary and composite fields, but it is curious 
to find phenomenologically viable solutions.

\section{Conclusion}

We have shown that the desirable features of the MSSM such as
a natural solution to the hierarchy problem, gauge coupling unification 
and a dark matter candidate can be incorporated into a partly supersymmetric 
grand unified model without inheriting the flavour, CP and gravitino/moduli 
problems characteristic of the MSSM. In the partly supersymmetric model all 
superpartners except those associated with the Higgs sector are superheavy. 
Grand unification is achieved not from the gauginos and Higgsinos, which
no longer contribute to the running, but instead from the partly elementary 
standard model matter which forms incomplete $SU(5)$ multiplets. Using a 
one-loop bulk calculation which corresponds to two-loops in the CFT we have 
shown that gauge coupling unification is achieved within $10\%$ of the 
experimental value. This was obtained for a two Higgs doublet model, where 
the $SU(5)$ gauge symmetry is broken by boundary conditions. The parities 
of the matter fields are then restricted by the $SU(5)$ breaking on the 
orbifold and unification is achieved with a composite $\tau_R,b_R$ and $t_R$.

It should be emphasized that the idea of having partly elementary standard 
model matter that contributes to the differential running of gauge couplings 
is quite general, and there exist other possibilities of grand unification 
with partly composite fermions. We briefly considered the first and second 
generations to be partly composite as well as choosing more general boundary 
conditions for the matter fields. In some cases this allowed the gauge 
coupling unification to become more precise at the leading log level and also 
allowed the unification scale to be near the Planck scale. This new class 
of models with part composite and part elementary fields contains many 
desirable features and offers new possibilities for gauge coupling 
unification.

\section*{Acknowledgments}
I would like to thank K. Agashe, N. Arkani-Hamed, K. Choi, A. Pomarol 
and R. Sundrum for useful discussions. It is a pleasure to acknowledge 
the hospitality of the CERN Theory Division and the Aspen Center for 
Physics where part of this work was done. This research was supported 
in part by a Department of Energy grant DE-FG02-94ER40823 at the University 
of Minnesota, a grant from the Office of the Dean of the Graduate School 
of the University of Minnesota, and an award from Research Corporation.


\begin{thebibliography}{99}

\bibitem{gp4} T.~Gherghetta and A.~Pomarol,
Phys.\ Rev.\ D {\bf 67}, 085018 (2003)
[arXiv:hep-ph/0302001].

\bibitem{rs}
L.~Randall and R.~Sundrum,
Phys.\ Rev.\ Lett.\  {\bf 83}, 3370 (1999)
[arXiv:hep-ph/9905221].

\bibitem{luty} 
M.~A.~Luty,
Phys.\ Rev.\ Lett.\  {\bf 89}, 141801 (2002)
[arXiv:hep-th/0205077].

\bibitem{adscft}
J.~M.~Maldacena,
Adv.\ Theor.\ Math.\ Phys.\  {\bf 2}, 231 (1998)
[arXiv:hep-th/9711200];
E.~Witten,
Adv.\ Theor.\ Math.\ Phys.\  {\bf 2}, 253 (1998)
[arXiv:hep-th/9802150];
S.~S.~Gubser, I.~R.~Klebanov and A.~M.~Polyakov,
Phys.\ Lett.\ B {\bf 428}, 105 (1998)
[arXiv:hep-th/9802109].

\bibitem{phenoadscft}
N.~Arkani-Hamed, M.~Porrati and L.~Randall,
JHEP {\bf 0108}, 017 (2001)
[arXiv:hep-th/0012148];
R.~Rattazzi and A.~Zaffaroni,
JHEP {\bf 0104}, 021 (2001)
[arXiv:hep-th/0012248];
M.~Perez-Victoria,
JHEP {\bf 0105}, 064 (2001)
[arXiv:hep-th/0105048].

\bibitem{log}
A.~Pomarol,
Phys.\ Rev.\ Lett.\  {\bf 85}, 4004 (2000)
[arXiv:hep-ph/0005293].

\bibitem{gr}
W.~D.~Goldberger and I.~Z.~Rothstein,
Phys.\ Rev.\ D {\bf 68}, 125011 (2003)
[arXiv:hep-th/0208060].

\bibitem{ads}
K.~Agashe, A.~Delgado and R.~Sundrum,
Annals Phys.\  {\bf 304}, 145 (2003)
[arXiv:hep-ph/0212028].

\bibitem{gp2}
T.~Gherghetta and A.~Pomarol,
Nucl.\ Phys.\ B {\bf 602}, 3 (2001)
[arXiv:hep-ph/0012378].

\bibitem{wgu}
W.~D.~Goldberger, Y.~Nomura and D.~R.~Smith,
Phys.\ Rev.\ D {\bf 67}, 075021 (2003)
[arXiv:hep-ph/0209158].

\bibitem{acs}
K.~Agashe, R.~Contino, and R.~Sundrum, to appear.

\bibitem{gp1}
T.~Gherghetta and A.~Pomarol,
Nucl.\ Phys.\ B {\bf 586}, 141 (2000)
[arXiv:hep-ph/0003129].

\bibitem{kawamura}
Y.~Kawamura,
Prog.\ Theor.\ Phys.\  {\bf 105}, 999 (2001)
[arXiv:hep-ph/0012125].


\bibitem{hn1}
L.~J.~Hall and Y.~Nomura,
Phys.\ Rev.\ D {\bf 64}, 055003 (2001)
[arXiv:hep-ph/0103125].

\bibitem{hmr}
A.~Hebecker and J.~March-Russell,
Nucl.\ Phys.\ B {\bf 613}, 3 (2001)
[arXiv:hep-ph/0106166].

\bibitem{bhn}
R.~Barbieri, L.~J.~Hall and Y.~Nomura,
Phys.\ Rev.\ D {\bf 66}, 045025 (2002)
[arXiv:hep-ph/0106190].

\bibitem{cp}
R.~Contino and A.~Pomarol,
arXiv:hep-th/0406257.

\bibitem{nda}
Z.~Chacko, M.~A.~Luty and E.~Ponton,
JHEP {\bf 0007}, 036 (2000)
[arXiv:hep-ph/9909248];
Y.~Nomura,
Phys.\ Rev.\ D {\bf 65}, 085036 (2002)
[arXiv:hep-ph/0108170].

\bibitem{bc}
R.~Contino, P.~Creminelli and E.~Trincherini,
JHEP {\bf 0210}, 029 (2002)
[arXiv:hep-th/0208002].



\bibitem{gr2}
W.~D.~Goldberger and I.~Z.~Rothstein,
Phys.\ Rev.\ D {\bf 68}, 125012 (2003)
[arXiv:hep-ph/0303158].

\bibitem{choi}
K.~w.~Choi and I.~W.~Kim,
Phys.\ Rev.\ D {\bf 67}, 045005 (2003)
[arXiv:hep-th/0208071].

\bibitem{adms}
K.~Agashe, A.~Delgado, M.~J.~May and R.~Sundrum,
JHEP {\bf 0308}, 050 (2003)
[arXiv:hep-ph/0308036].

\end{thebibliography}
\end{document}